\documentclass[a4paper,11pt]{article}
\usepackage{nlc2}

\title{NLC-2 graph recognition and isomorphism} 

\author{Vincent Limouzy$^1$ \and Fabien de Montgolfier$^1$  \and Micha\"el Rao$^1$ }

\date{}

\begin{document}

\maketitle
\setcounter{footnote}{1}
\footnotetext{LIAFA, Université Paris 7. \texttt{\{limouzy,fm,rao\}@liafa.jussieu.fr}. Research supported by the French ANR project ``Graph Decompositions and Algorithms (GRAAL)''}

\begin{abstract}
NLC-width is a variant of clique-width with many application in graph
algorithmic. This paper is devoted to graphs of NLC-width 
two. After giving new structural properties of the class, we propose a
$O(n^2 m)$-time algorithm, improving 
Johansson's algorithm~\cite{Johansson00}. 
Moreover, our alogrithm is simple to understand.
The above properties and
algorithm allow us to propose a robust $O(n^2 m)$-time isomorphism
algorithm for NLC-2 graphs. As far as we know, it is the first
polynomial-time algorithm.
\end{abstract}

\section{Introduction}
NLC-width is a graph parameter introduced by Wanke \cite{Wanke94}. This
notion is tightly related to clique-width introduced by Courcelle
\emph{et al.}~\cite{CourcelleER93}. Both parameters were introduced to generalise the well
known tree-width. The motivation on research about such \emph{width}
parameter is that, when the width (NLC-, clique- or tree-width) %whose definition depends on the
is bounded by a constant, then many NP-complete
problems can be solved in polynomial  (even linear) time, if the
decomposition is provided.

Such parameters give insights on graph structural properties. 
Unfortunately, finding the minimum NLC-width of the graph was shown to be 
NP-hard by Gurski \emph{et al.} \cite{GurskiW05}. Some results however are known. Let NLC-$k$ be the class of graph of NLC width bounded by $k$. NLC-1 is  exactly the class of cographs. Probe-cographs, bi-cographs and weak-bisplit graphs~\cite{Fouquet99} belong to NLC-2. 
Johansson \cite{Johansson00} proved that recognising NLC-2 graphs is polynomial and provided an $O(n^{4} \log(n))$ recognition algorithm. Complexity for recognition of NLC-$k$, $k\ge 3$, is still unknown.

In this paper we improve Johansson's result down to $O(n^{2} m)$. Our
approach relies on graph decompositions.  We establish the tight links
that exist between NLC-2 graphs and the so-called modular decomposition,
split decomposition, and bi-join decomposition.

NLC-2 can be defined as a graph colouring problem.  Unlike NLC-$k$
classes, for $k \ge 3$, \emph{recolouring} is useless for prime NLC-2 graphs. That allow us to
propose a canonical decomposition of bi-coloured NLC-2 graphs, defined
as certain bi-coloured split operations. This
decomposition can be computed in $O(nm)$ time if the colouring is
provided. If a graph is \emph{prime}, there using split and bi-join 
decompositions, we show that there is at most $O(n)$
colourings to check. Finally, modular decomposition properties allow
to reduce NLC-2 graph decomposition to prime NLC-2 graph
decomposition. Section~\ref{sectreco} explains this $O(n^2m)$-time
decomposition algorithm.

In Section~\ref{sectiso} is proposed an isomorphism algorithm. Using modular, split and bi-join decompositions and the canonical NLC-2 decomposition, isomorphism between two NLC-2 graphs can be tested in $O(n^2m)$ time.

\section{Preliminaries}

A graph $G=(V,E)$ is pair of a set of \emph{vertices} $V$ and a set of \emph{edges} $E$. 
For a graph $G$, $V(G)$ denote its set of vertices, $E(G)$ its set of edges, $n(G)=\vert V(G) \vert$ and $m(G) = \vert E(G) \vert$ (or $V$, $E$, $n$ and $m$ if the graph is clear in the context).
$N(x)= \{y \in V : \{x,y\}\in E\}$  denotes the \emph{neighbourhood} of the vertex $x$, and $N[x] = N(v) \cup \{v\}$. 
For $W\subseteq V$, $G[W]= (W, E \cap W^2)$ denote the \emph{graph induced by $W$}.
Let $A$ and $B$ be two disjoint subsets of $V$. Then we note $A\join B$ if for all $(a,b)\in A\times B$, then $\{a,b\} \in E$, and we note $A\disjoin B$ if for all $(a,b)\in A\times B$, then $\{a,b\} \not\in E$.
Two graphs $G=(V,E)$ and $G'=(V',E')$ are \emph{isomorphic} (noted $G \iso G'$)
if there is a bijection $\varphi: V \rightarrow V' $ such that 
$\{x,y\} \in E \Leftrightarrow \{\varphi(x),\varphi(y)\} \in E'$, for all $u,v\in V$.

A \emph{$k$-labelling} (or \emph{labelling}) is a function $l : V \to \{1,\ldots, k\}$. 
A \emph{$k$-labelled graph} is a pair of a graph $G=(V,E)$ and a $k$-labelling $l$ on $V$. It is denoted by $(G,l)$ or by $(V,E,l)$. 
Two labelled graphs $(V,E,l)$ and $(V',E',l')$ are isomorphic if there is a bijection $\varphi:V\to V'$ such that $\{u,v\}\in E \Leftrightarrow \{\varphi(x),\varphi(y)\} \in E'$ and $l(u)=l'(\varphi(u))$ for all $u,v \in V$.

\paragraph{NLC-$k$ classes.}

Let $k$ be a positive integer. The class of NLC-$k$ graphs is defined recursively by the following operations.

\begin{itemize}
\item For all $i \in \entkk$, $\vlab{i}$ is in NLC-$k$, where $\vlab{i}$ is the graph with one vertex labelled $i$.
\item Let $G_1=(V_1,E_1,\lab_1)$ and $G_2=(V_2,E_2,\lab_2)$ be NLC-$k$ and let $S\subseteq {\entkk}^2$. Then $G_1 \nlcop{S} G_2$ is in NLC-$k$, where
\vspace{-0.1cm}
$G_1 \nlcop{S} G_2 = (V,E,\lab)$ with $V=V_1\cup V_2$,
$$E=E_1 \cup E_2 \cup \{\{u,v\} : u\in V_1, v\in V_2, (\lab_1(u),\lab_2(v)) \in S\}$$ 
\vspace{-.4cm}
$$\text{and for all $u\in V$, }\lab(u) = \begin{cases} \lab_1(u) \text{ if $u\in V_1$}\\ \lab_2(u) \text{ if $u\in V_2$.}\end{cases}$$ 
\item Let $R: \entkk \rightarrow \entkk$ and $G=(V,E,\lab)$ be NLC-$k$. Then $\rho_{R}(G)$ is in NLC-$k$, where $\rho_{R}(G)=(V,E,\lab')$ such that $\lab'(u)=R(\lab(u))$ for all $u\in V$.
\end{itemize}
A graph is NLC-$k$ if there is a $k$-labelling of $G$ such that $(G,l)$ is in NLC-$k$. 
A $k$-labelled graph is \emph{NLC-$k$ $\rho$-free} if it can be constructed without the $\rho_R$ operation.

\paragraph{Modules and modular decomposition.}
A \emph{module} in a graph is a non-empty subset $X\subseteq V$ such that for all $u\in V \setminus X$, 
then either $N(u)\cap X = \emptyset$ or $X\subseteq N(u)$.
A module is \emph{trivial} if $\vert X\vert \in \{1,\vert V \vert \}$. A graph is \emph{prime} (w.r.t. modular decomposition) if all its modules are trivial.
Two sets $X$ and $X'$ \emph{overlap} if $X\cap X'$,$X\setminus X'$ and $X' \setminus X$ are non-empty. 
A module $X$ is \emph{strong} if there is no module $X'$ such that $X$ and $X'$ overlap. 
Let $\mm'(G)$ be the set of modules, let $\mm(G)$  be the set of strong modules of $G$, and let $\mp(G)=\{M_1,\ldots,M_k\}$ be the maximal (w.r.t. inclusion) members of $\mm(G) \setminus \{V\}$.

\begin{theorem}\cite{Gallai67}
Let $G=(V,E)$ be a graph such that $\vert V\vert \ge 2$. Then: 
\begin{itemize}
\item if $G$ is not connected, then $\mp(G)$ is the set of connected components of $G$,
\item if $\overline{G}$ is not connected, then $\mp(G)$ is the set of connected components of $\overline{G}$,
\item if $G$ and $\overline{G}$ are connected, then $\mp(G)$ is a partition of $V$ and is formed with the maximal members of $\mm'\setminus \{V\}$.
\end{itemize}
\end{theorem}

In all cases, $\mp(G)$ is a partition of $V$, and $G$ can be
decomposed into $G[M_1],\ldots,G[M_k]$.
The \emph{characteristic graph} $G^*$ of a graph $G$ is the graph of vertex set $\mp(G)$ and two $P,P'\in \mp(G)$ are adjacent if there is an edge between $P$ and $P'$ in $G$ (and so there is no non-edges since $P$ and $P'$ are two modules).
The recursive decomposition of a graph by this operation gives the \emph{modular decomposition} of the graph, and can be represented by a rooted tree, called the \emph{modular decomposition tree}. It can be computed in linear time~\cite{McConnell99}.
The nodes of the modular decomposition tree are exactly the strong modules, so in the following we make no distinction between the modular decomposition of $G$ and $\mm(G)$. 
Note that $\vert \mm(G) \vert \le 2 \times n - 1$. 
For $M\in \mm(G)$, let $G_M=G[M]$ and $G^*_M$ its characteristic graph.

\begin{lemma}\cite{Johansson00}
\label{lem:nlcprime}
Let $G$ be a graph.  $G$ is NLC-$k$ if and only if every characteristic graph in the modular decomposition of $G$ is NLC-$k$.
\end{lemma}

Moreover, a NLC-$k$ expression for $G$ can be easily constructed from the modular decomposition and from NLC-$k$ expressions of prime graphs. On prime graphs, NLC-2 recognition is easier:

\begin{lemma}\cite{Johansson00}
\label{lem:nlcrfprime}
Let $G$ be a prime graph. Then $G$ is NLC-2 if and only if there is a $2$-labelling $l$ such that $(G,l)$ is \nlcrf.
\end{lemma}

\paragraph{Bi-partitive family.}
A \emph{bipartition} of $V$ is a pair $\{X,Y\}$ such that $X\cap Y = \emptyset$, $X\cup Y=V$ and $X$ and $Y$ are both non empty.
Two bipartitions $\{X,Y\}$ and $\{X',Y'\}$ \emph{overlap} if $X\cap Y$, $X\cap Y'$, $X'\cap Y$ and $X'\cap Y'$ are non empty.
A family $\mf$ of bipartitions of $V$ is \emph{bipartitive} if 
(1) for all $v\in V$, $\{\{v\},V\setminus \{v\}\} \in \mf$ and
(2) for all $\{X,Y\}$ and $\{X',Y'\}$ in $\mf$ such that $\{X,Y\}$ and $\{X',Y'\}$ overlap, then $\{X\cap X', Y\cup Y'\}$, $\{X\cap Y', Y\cup X'\}$, $\{Y\cap X', X\cup Y'\}$, $\{Y\cap Y', X\cup X'\}$ and $\{X\Delta X', X\Delta Y'\}$ are in $\mf$ (where $X\Delta Y = (X\setminus Y) \cup (Y\setminus X)$).
Bipartitive families are very close to partitive families~\cite{CHM81}, which generalise properties of modules in a graph.

A member $\{X,Y\}$ of a bipartitive family $\mf$ is \emph{strong} if there is no $\{X',Y'\}$ such that $\{X,Y\}$ and $\{X',Y'\}$ overlap.
Let $T$ be a tree. For an edge $e$ in the tree, $\{\compe{1},\compe{2}\}$ denote the bipartition of leaves of $T$ such that two leaves are in the same set if and only if the path between them avoids $e$.
Similarly, for an internal node $\alpha$, $\{\compa{1},\ldots, \compa{d(\alpha)}\}$ denote the partition of leaves of $T$ such that two leaves are in the same set if and only if the path between them avoid $\alpha$.

\begin{theorem}\cite{CunninghamE80}
Let $\mf$ be a bipartitive family on $V$. Then there is an unique
unrooted tree $T$, called the \emph{representative tree} of $\mf$,
such that the set of leaves of $T$ is $V$, the internal nodes of $T$
are labelled \texttt{degenerate} or \texttt{prime}, and\\ \indent
~~~-~ for every edge $e$ of $T$, $\{\compe{1},\compe{2}\}$ is a strong
member of $\mf$, and there is no other strong member in $\mf$,\\
\indent ~~~-~ for every node $\alpha$ labelled \texttt{degenerate},
and for every $\emptyset \subsetneq I \subsetneq \{1,\ldots,
d(\alpha)\}$,\\ $\{\cup_{i\in I} \compa{i}, V \setminus \cup_{i\in I}
\compa{i}\}$ is in $\mf$, and
there is no other member in $\mf$.
\end{theorem}

\paragraph{Split decomposition.}
A \emph{split} in a graph $G=(V,E)$ is a bipartition $\{X,Y\}$ of $V$ such that the set of vertices in $X$ having a neighbour in $Y$ have the same neighbourhood in $Y$ (\emph{i.e.}, for all $u,v\in X$ such that $N(u)\cap Y\ne \emptyset$ and $N(v)\cap Y\ne \emptyset$, then $N(u)\cap Y = N(v)\cap Y$). 
A \emph{co-split} in a graph $G$ is a split in $\overline{G}$.
The family of split in a connected graph is a bipartitive family~\cite{Cunningham82}. The split decomposition tree is the representative tree of the family of splits, and can be computed in linear time~\cite{Dahlhaus00}.
Let $\alpha$ be an internal node of the split decomposition tree of a connected graph $G$. For all $i\in \{1,\ldots, d(\alpha)\}$ let $v_i \in \compa{i}$ such that $N(v_i)\setminus \compa{i} \ne \emptyset$. Since $G$ is connected, such a $v_i$ always exists. $G[\{v_1,\ldots, v_{d(\alpha)}\}]$ denote the \emph{characteristic graph} of $\alpha$. The characteristic graph of a \texttt{degenerate} node is a complete graph or a star~\cite{Cunningham82}.

\paragraph{Bi-join decomposition.}
A \emph{bi-join} in a graph is a bipartition $\{X,Y\}$ such that for all $u,v\in X$, $\{N(u)\cap Y,Y\setminus N(u)\} = \{N(v)\cap Y,Y\setminus N(v)\}$. The family of bi-joins in a graph is bipartitive. The \emph{bi-join decomposition tree} is the representative tree of the family of bi-joins, and can be computed in linear time~\cite{DeMontgolfieR05,DeMontgolfieR05b}.
Let $\alpha$ be an internal node of the bi-join decomposition tree of a graph $G$. For all $i\in \{1,\ldots, d(\alpha)\}$ let $v_i \in \compa{i}$. $G[\{v_1,\ldots, v_{d(\alpha)}\}]$ denote the \emph{characteristic graph} of $\alpha$. 
The characteristic graph of a \texttt{degenerate}  node is a complete bipartite graph or a disjoint union of two complete graphs~\cite{DeMontgolfieR05,DeMontgolfieR05b}.

\begin{figure}
 \begin{center}
  \includegraphics[scale=1]{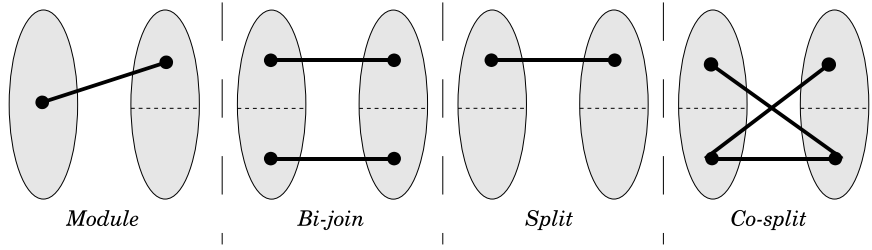}
 \end{center}
\caption{A module, a bi-join, a split and a co-split}
\label{fig:ModSplitCoSplit}
\end{figure}

\section{Recognition of NLC-2 graphs}\label{sectreco}

\subsection{\nlcrf canonical decomposition}

In this section, $G=(V,E,l)$ is a 2-labelled graph such that every mono-coloured module (\emph{i.e.} a module $M$ such that $\forall v,v' \in M$, $l(v)=l(v')$) has size $1$.
A couple $(X,Y)$ is a \emph{\cutnox} if $X\cup Y = V$, $X\cap Y= \emptyset$, $X\ne \emptyset$ and $Y\ne \emptyset$.
Let $S\subseteq \bs$. A \cut $(X,Y)$ is a \emph{\scutnox} of $G$ if for all $u\in X$ and $v\in Y$, then $\{u,v\} \in E$ if and only if $(l(u),l(v)) \in S$.
For $S\subseteq \bs$ let $\mf_S(G)$ be the set of \scut of $G$.

\begin{definition}[Symmetry]
\label{def:symmetry}
We say that $S\in \bs$ is \emph{symmetric} if $(1,2) \in S \iff (2,1) \in S$, otherwise we say that $S$ is \emph{non-symmetric}.
\end{definition}
\begin{definition}[Degenerate property]
\label{def:degenerate}
A family $\mf$ of \cuts has the \emph{degenerate property} if there is a partition $\mp$ of $V$ such that for all $\emptyset \subsetneq \mathcal{X} \subsetneq \mp$, $(\bigcup_{X\in \mathcal{X}} X, \bigcup_{Y\in \mp \setminus \mathcal{X}} Y)$ is in $\mf$, and there is no others \cut in $\mf$.
\end{definition}

\begin{lemma}\label{lem:syminter}
For every symmetric $S \subseteq \bs$,
$\mf_S(G)$ has the degenerate property.
\end{lemma}
\begin{proof}
The family $\mf_{\{\}}(G)$ has the degenerate property since $(X,Y)$ is a \Scut{\{\}} if and only if there is no edges between $X$ and $Y$ ($\mp$ is exactly the connected components).
For $W\subseteq V$, let $\lc{G}{W}= (V,E \Delta W^2,l)$. For $i\in \{1,2\}$ let $V_i = \{ v\in V : l(v)=i \}$.
Let $G_1=\lc{G}{\lv{1}}$, $G_2=\lc{G}{\lv{2}}$ and $G_{12}=\lc{(\lc{G}{\lv{1}})}{\lv{2}}$.
\begin{itemize}
\item $\mf_{\{(1,1)\}}(G)=\mf_{\{\}}(G_1)$,
$\mf_{\{(2,2)\}}(G)=\mf_{\{\}}(G_2)$,
$\mf_{\{(1,1),(2,2)\}}(G)=\mf_{\{\}}(G_{12})$,
\item $\mf_{\{(1,1),(1,2),(2,1),(2,2)\}}(G)=\mf_{\{\}}(\overline{G})$,
$\mf_{\{(1,2),(2,1),(2,2)\}}(G)=\mf_{\{\}}(\overline{G_1})$,\\
$\mf_{\{(1,1),(1,2),(2,1)\}}(G)=\mf_{\{\}}(\overline{G_2})$,
$\mf_{\{(1,2),(2,1)\}}(G)=\mf_{\{\}}(\overline{G_{12}})$.
\end{itemize}
Thus for every symmetric $S \subseteq \bs$, $\mf_S(G)$ has the degenerate property.\end{proof}

\begin{definition}[Linear property]
\label{def:linear}
A family $\mf$ of \cuts has the \emph{linear property} if for all $(X,Y)$ and $(X',Y')$ in  $\mf$, 
either $X\subseteq X'$ or $X'\subseteq X$. 
\end{definition}

\begin{lemma}
For every non-symmetric $S \subseteq \bs$,
$\mf_S(G)$ has the linear property.
\end{lemma}
\begin{proof}
Case $S=\{(1,2)\}$: suppose that $X\setminus X'$ and $X'\setminus X$ are both non-empty. Then if $u\in X\setminus X'$ is labelled $1$ and $v\in X'\setminus X$ is labelled $2$, $u$ and $v$ has to be adjacent and non-adjacent, contradiction. Thus $X\setminus X'$ and $X'\setminus X$ are mono-coloured.
Now suppose w.l.o.g. that all vertices in $X\Delta X'$ are labelled $1$. Then $X\Delta X'$ is adjacent to all vertices labelled $2$ in $Y\cap Y'$ and non adjacent to all vertices labelled $1$ in $Y\cap Y'$. Moreover $X\Delta X'$ is non adjacent to all vertices in $X\cap X'$. Thus $X\Delta X'$ is a mono-coloured module, and $\vert X\Delta X' \vert \ge 2$. Contradiction.
For others non-symmetric $S$, we bring back to case $\{(1,2)\}$ like in the proof of lemma~\ref{lem:syminter}.
\end{proof}

For $S\subseteq \bs$, let $\mp_S(G)$ denote the unique partition of $V$ such that 
(1) for all $(X,Y) \in \mf_{S}(G)$ and $P\in \mp_S(G)$, $P\subseteq X$ or $P\subseteq Y$, and 
(2) for all $P,P' \in \mp$, $P\ne P'$, there is a $(X,Y) \in \mf_{S}(G)$ such that $P\subseteq X$ and $P'\subseteq Y$, or $P\subseteq Y$ and $P'\subseteq X$.
For a non-symmetric $S\in \bs$, let $\mp'_S(G) = (P_1, \ldots , P_k)$ denote the unique ordering of elements in $\mp_S(G)$ such that for all $(X,Y)\in \mf_S(G)$, there is a $l$ such that $X=\cup_{i\in \{1,\ldots , l\}} P_i$.

\begin{lemma}\label{lem:decomptree}
If $G$ is in \nlcrf, then there is a $S \subseteq \bs$ such that $\mf_S(G)$ is non-empty.
\end{lemma}
\begin{proof}
If $G$ is \nlcrf, then there is a $S\subseteq \bs$, and two graphs $G_1$ and $G_2$ such that $G=G_1 \times_S G_2$. Thus $(V(G_1),V(G_2)) \in \mf_S(G)$ and $\mf_S(G)$ is non empty. 
\end{proof}

\begin{lemma}\label{lem:decomptree3}
Let $G=(V,E,l)$ 2-labelled graph and let $S\subseteq \bs$.
If $G$ is \nlcrf and has no mono-coloured non-trivial module, then for all $P\in \mp_S(G)$, $G[P]$ has no mono-coloured non-trivial module.
\end{lemma}
\begin{proof}
If $M$ is a mono-coloured module of $G[P]$, then $M$ is a mono-coloured module of $G$. Contradiction.
\end{proof}

\begin{lemma}\label{lem:decomptree2}
Let $G=(V,E,l)$ 2-labelled graph and let $S\subseteq \bs$.
Then $G$ is \nlcrf if and only if for all $P\in \mp_S(G)$, $G[P]$ is \nlcrf.
\end{lemma}
\begin{proof}
The ``only if'' is immediate.
Now suppose that for all $P\in \mp_S(G)$, $G[P]$ is \nlcrf. 
If $S$ is symmetric, let $\mp_S(G)= \{P_1,\ldots, P_{\vert \mp_S(G)\vert}\}$. Then $G = ((G[P_1] \times_S G[P_2] )\times_S \ldots \times_S G[P_{\vert \mp_S(G)\vert}]$, and $G$ is \nlcrf.
Otherwise, if $S$ is non-symmetric, let $\mp'_S(G)= (P_1,\ldots, P_{\vert \mp_S(G)\vert})$. Then $G = ((G[P_1] \times_S G[P_2] )\times_S \ldots \times_S G[P_{\vert \mp_S(G)\vert}]$, and $G$ is \nlcrf.
\end{proof}

The \emph{\nlcrf decomposition tree} of a $2$-labelled graph $G$ is a
rooted tree such that the leaves are the vertices of $G$, and the
internal nodes are labelled by $\times_S$, with $S\subseteq \bs$. An
internal node is \texttt{degenerated} if $S$ is symmetric, and
\texttt{linear} if $S$ is non-symmetric.
By lemmas~\ref{lem:decomptree},~\ref{lem:decomptree3} and~\ref{lem:decomptree2}, $G$ is \nlcrf if and only if it has a \nlcrf decomposition tree.
This decomposition tree is not unique. But we can define a \emph{canonical decomposition tree} if we fix a total order on the subsets of $\bs$ (for example, the lexicographic order).
If two graphs are isomorphic, then they have the same canonical decomposition tree.
Algorithm~\ref{algo:treecomp} computes the canonical decomposition tree of a $2$-labelled prime graph, or fails if $G$ is not \nlcrf.

\begin{figure}[t]
\linesnumbered
\begin{algorithm}[H]
\SetKw{KwFail}{fail with}
\KwIn{A $2$-labelled graph $G=(V,E,l)$} 
\KwOut{A \nlcrf decomposition tree, or fail if $G$ is not \nlcrf}
\dontprintsemicolon 
\lIf{$\vert V \vert = 1$}{\Return{the leaf $\vlab{l(v)}$} (where $V = \{ v\}$)} \;
Let $\ms$ be the set of subsets of $\bs$ and $\sigma$ be the lexicographic order of $\ms$\;
\ForEach{ $S \in \ms$ w.r.t. $\sigma$ }{
Compute $\mp_S(G)$, and $\mp'_S(G)$ if $S$ is non-symmetric (see algorithm~\ref{algo:mpcomp})\; 
\If{$\vert \mp_S(G) \vert > 1$}{
  Create a new $\times_S$ node $\beta$ \;
  \ForEach{$P \in \mp_S(G)$ (w.r.t. $\mp'_S(G)$ if $S$ is non-symmetric)}{
    make \nlcrf decomposition tree of $G[P]$ be a child of $\beta$.
  }
  \Return{the tree rooted at $\beta$}
}
}
\KwFail{Not \nlcrf}
\caption{Computation of the \nlcrf canonical decomposition tree}
\label{algo:treecomp}
\end{algorithm}
\linesnotnumbered
\end{figure}

Algorithm~\ref{algo:mpcomp} computes $\mp_S$ and $\mp'_S$ for a $2$-labelled prime graph $G$ and $S\subseteq \bs$ in linear time.
We need some additional definitions for this algorithm and its proof of correctness.
A \emph{bipartite graph} is a triplet $(X,Y,E)$ such that $E \subseteq X \times Y$. The \emph{bi-complement} of a bipartite graph $(X,Y,E)$ is the bipartite graph $(X,Y,(X\times Y) \setminus E)$.
A \emph{bipartite trigraph (BT)} is a bipartite graph with two types of edges: the \emph{join} edges and the \emph{mixed} edges. It is denoted by $\btcal = \biptri$ where $E_j$ are the set of \emph{join} edges, and $E_m$ the set of \emph{mixed} edges. 
A \emph{\btmodulenox} in a \bt is a $M\subseteq X$ or $M\subseteq Y$ such that $M$ is a module in $(X,Y,E_j)$ and there is no \emph{mixed} edges between $M$ and $(X\cup Y) \setminus M$.
For $v\in X\cup Y$, let $N_j(v) = \{u\in X\cup Y : \{u,v\} \in E_j \}$ and 
$N_m(v) = \{u\in X\cup Y : \{u,v\} \in E_m \}$.
Let $d_j(v)= \vert N_j(v) \vert$ and $d_m(v) = \vert N_m(v) \vert$.
A \emph{\trucnox} in a \bt $\biptri$ is a \cut $(A,B)$ of $X \cup Y$, such that  there is no edges between $A\cap Y$ and $B\cap X$, and there is only \emph{join} edges between $A\cap X$ and $B\cap Y$.

In algorithm \ref{algo:mpcomp}, $\btcal$ is obtained from the graph $G$. Vertices of $X$ correspond to subsets of vertices labelled $1$ in $G$, and vertices of $Y$ correspond to subsets of vertices labelled $2$.
There is a \emph{join} edge between $M$ and $M'$ in $\btcal$ if $M\join M'$ in $G$,
and there is a \emph{mixed} edge between $M \in X$ and $M'\in Y$ in $\btcal$ if there is at least an edge and a non-edge between $M$ and $M'$ in $G$.
%% est-ce que c'est assez clair ?
Such a graph $\btcal$ can easily be built in linear time from a given graph $G$. It
suffices to consider a list and an array bounded by the number of component in $G$ with 
the same colour.
The following lemmas are close to observations in~\cite{Fouquet99}, but deal with \bts instead of bipartite graphs (proofs are given in appendix).

\begin{lemma}\label{lem:vanherpe}
Let $G = \biptri$ be a \bt such that every \btmodule has size $1$.
Let $(x_1,\ldots, x_{\vert X\vert})$ be $X$  sorted by $(d_j(x),d_m(x))$ in lexicographic decreasing order.
If $(A,B)$ is a \truc of $G$, then there is a $k\in \{0,\ldots , \vert X \vert\}$ such that $A \cap X =\{x_1,\ldots ,x_k\}$.
\end{lemma}
\begin{comment}
\begin{proof}
For all $v\in A\cap X$, $d_j(v) \ge \vert B\cap Y \vert$, and 
for all $v\in B\cap X$, $d_j(v) \le \vert B\cap Y \vert$.
Moreover, if there is a $v\in B\cap X$ with $d_j(v) = \vert B\cap Y \vert$, then $d_m(v)=0$.
Let $C= \{ v\in X  : d_j(v) = \vert B\cap Y \vert \text{ and } d_m(v) = 0\}$.
Then $C$ is a \btmodule of $G$, and thus $\vert C \vert \le 1$.
Every vertex in $A\cap X\setminus C $ are before every vertex in $B\cap X \setminus C$ in the ordering. 
Moreover, if $\vert C \vert >0$, then vertices in $A\cap X\setminus C$ are before the vertex in $C$, and vertices in $B\cap X\setminus C$ are after the vertex in $C$ in the ordering.
\end{proof}
\end{comment}

\begin{lemma}\label{lem:vanherpe2}
Let $k\in \{0,\ldots, \vert X \vert\}$ and $k'\in \{0,\ldots, \vert Y \vert\}$.
Then $(A,(X\cup Y) \setminus A)$,
where $A=\{x_1,\ldots, x_k, y_1,\ldots, y_{k'} \}$,
is a \truc of $G$  if and only if %there is a $k\in \{0,\ldots\, \vert X \vert\}$ and $k'\in \{0,\ldots\, \vert Y \vert\}$ such that 
$\sum_{i=1}^k d_j(x_i) - \sum_{i=1}^{k'} d_j(y_i) = k \times ( \vert Y \vert - k')$ and $\sum_{i=1}^k d_m(x_i) - \sum_{i=1}^{k'} d_m(y_i) = 0$.
\end{lemma}
\begin{comment}
\begin{proof}
The ``If'' part is by definition. Now let us consider the ``Only if''
part. Let us assume that the degree condition holds. %, but $(A,B)$ is
not a semi-join.  We will denote $a$ the number of join edges between
$A\cap X$ and $B \cap Y$, $b$ the number of join edges between $A \cap
X$ and $A\cap Y$, and $c$ the number of mixed edges between $A \cap X$
and $A\cap Y$.  Note that $a \le k (\vert Y\vert - k')$, and $b\le
\sum_{i=1}^{k'} d_j(y_i)$, thus $a \ge k (\vert Y\vert - k')$. So we
have $a = k (\vert Y\vert - k')$, and $\sum_{i=1}^{k'} d_j(y_i) - b
=0$. In other words, there are only join edges between $A\cap X$ and
$B \cap Y$, and there is no join edge between $A\cap Y$ and $B\cap X$.
Now since there is only join edges between $A\cap X$ and $B \cap Y$, $
c = \sum_{i=1}^k d_m(x_i) = \sum_{i=1}^{k'} d_m(y_i)$, thus there is
no mixed edges between $A\cap Y$ and $B\cap X$.
\end{proof}
\end{comment}

\linesnumbered
\begin{algorithm}[t]
\SetKw{KwOr}{or}
\SetKw{KwAnd}{and}
\KwIn{A $2$-labelled graph $G$, and $S\subseteq\bs$} 
\KwOut{$\mp_S$ if $S$ is symmetric, $\mp'_S$ if $S$ is non-symmetric}
\SetKw{Vi}{$V_{i}$}
\Vi $\gets \{v : v \in V$ and $l(v)=i\}$ \;
\lIf{$(1,1) \in S$}{
  \nllabel{li:a} $\setmod_1 \gets$ co-connected components of $G[V_1]$\; 
}
\lElse{
  \nllabel{li:b} $\setmod_1  \gets$ connected components of $G[V_1]$\; 
}
%%%%%%%%
\lIf{$(2,2) \in S$}{
  \nllabel{li:c}	
  $\setmod_2 \gets$ co-connected components of $G[V_2]$\; 
}
\lElse{
  \nllabel{li:d} 
  $\setmod_2  \gets$ connected components of $G[V_2]$\; 
}

\nllabel{li:biparti}$\mathcal{B}=(\setmod_1,\setmod_2,E_j,E_m)$ $\gets$ 
the bipartite trigraph between the elements of $\setmod_1$ and $\setmod_2$\ ;

\If{$S \cap \{(1,2),(2,1)\} = \emptyset$}{
  
  \Return{\nllabel{li:e} connected components of $(\setmod_1,\setmod_2,E_j \cup E_m)$}
}
\ElseIf{$S \cap \{(1,2),(2,1)\} = \{(1,2),(2,1)\}$}{
  
  \Return{\nllabel{li:f} connected components of the bi-complement of $(\setmod_1,\setmod_2,E_j)$}}
\lElse{\nllabel{li:algotruc} Search all \trucs of $\btcal$ (see appendix) \;
}
\caption{Computation of $\mp_S$ and $\mp'_S$}
\label{algo:mpcomp}
\end{algorithm}
\linesnotnumbered

\begin{theorem}
\label{th:correctness_mpcomp}
Algorithm~\ref{algo:mpcomp} is correct and runs in linear time.
\end{theorem}
\begin{proof}
{\bf Correctness:}
Suppose that $(A,B)$ is a \scut. If $(1,1) \not\in S$, then there is no edge between $A\cap V_1$ and $B\cap V_1$, thus $(A,B)$ cannot cut a component $\setmod_1$ (and similarly for $(1,1) \in S$, and for $\setmod_2$).
Now we work on the \bt $\mathcal{B}=(\setmod_1,\setmod_2,E_j,E_m)$. If $S\cap \{(1,2),(2,1)\} = \emptyset$, then \scuts correspond exactly to connected components of $\mathcal{B}$, and if $S\cap \{(1,2),(2,1)\} = \{(1,2),(2,1)\}$ then \scuts correspond exactly to connected components of the \bt of $\overline{G}$, which is $(\setmod_1,\setmod_2,(\setmod_1 \times \setmod_2) \setminus (E_j \cup E_m),E_m)$. Finally, if $S$ is non-symmetric, \scuts correspond to \trucs of $\mathcal{B}$ (see appendix).

{\bf Complexity}:
It is well admitted that we can perform a BFS on a graph or its complement in linear time~\cite{HabibPV99,DahlhausGM02}.
The instructions on lines [\ref{li:a}%, \ref{li:b}, \ref{li:c}, 
-\ref{li:d},\ref{li:e}] can be done with a BFS on a graph or its complement. 
It is easy to see that we can do a BFS on the bi-complement in linear time (like a BFS on a complement graph, with two vertex lists for $X$ and $Y$), so instruction line \ref{li:f} can be done in linear time.
Finally, the operations at line \ref{li:algotruc} are done in linear time (see appendix).
\end{proof}

These results can be summarized as:
\begin{theorem}
Algorithm~\ref{algo:treecomp} computes the canonical \nlcrf
decomposition tree of a 2-labelled graph in $O(nm)$ time.
\end{theorem}

\subsection{NLC-2 decomposition of a prime graph}

In this section, $G$ is an unlabelled prime (w.r.t. modular decomposition) graph, with $\vert V\vert \ge 3$.

\begin{definition}[\emph{$2$-bimodule}]
A bipartition $\{X,Y\}$ of $V$ is a \emph{$2$-bimodule} if $X$ can be partitioned into $X_1$ and $X_2$, and $Y$ into $Y_1$ and $Y_2$ such that for all $(i,j)\in \bs$, then either $X_i \disjoin Y_j$ or $X_i \join Y_j$. 
It is easy to see that if $\{X,Y\}$ is a $2$-bimodule if and only if $\{X,Y\}$ is a split, a co-split or a bi-join.
Moreover, if $\min (\vert X \vert,\vert Y \vert) > 1$ then $\{X,Y\}$ cannot be 
both of them in the same time (since $G$ is prime).
\end{definition}
Let $l : V \to \{1,2\}$ be a 2-labelling. Then $s(l)$ denote the 2-labelling on $V$ such that for all $v\in V$, $s(l)(v) = 1$ if and only if $l(v)=2$.
\begin{definition}[Labelling induced by a $2$-bimodule]
Let $\{X,Y\}$ be a \proper $2$-bimodule. We define the \emph{labelling $l:V \to \{1,2\}$ of $G$ induced by $\{X,Y\}$}. 
If $\vert X \vert = \vert Y\vert =1$, then $l(x)=1$ and $l(y)=2$, where $X = \{x\}$ and $Y = \{y\}$.
If $\vert X \vert = 1$, then $l(v)=1$ iff $v\in N[x]$. 
Similarly if $\vert Y \vert = 1$, then $l(v)=1$ iff $v\in N[y]$. 
Now we suppose $\min(\vert X \vert , \vert Y \vert) > 1$.
If $\{X,Y\}$ is a split, then the set of vertices in $X$ with a neighbour $Y$ and the set of vertices in $Y$ with a neighbour in $X$ is labelled $1$, others vertices are labelled $2$.
If $\{X,Y\}$ is a co-split, then a labelling of $G$ induced by $\{X,Y\}$ is a labelling of $\overline{G}$ induced by the split $\{X,Y\}$.
Finally if $\{X,Y\}$ is a bi-join, $l$ is such that $\{v \in X: l(v)=1 \}$ is a join with $\{ v \in Y: l(v)=1 \}$ and $\{v \in X: l(v)=2 \}$ is a join with $\{ v \in Y: l(v)=2 \}$. Note that if $\{X,Y\}$ is a bi-join, then there is two possibles labelling $l_1$ and $l_2$, with $l_1=s(l_2)$.
If $\{X,Y\}$ is a $2$-bimodule of $G$ and $l$ a labelling induced by $\{X,Y\}$, then every mono-coloured module has size $1$ (since $G$ is prime and $\vert V \vert \ge 3$).
\end{definition}

\begin{definition}[Good $2$-bimodule]
A \proper $2$-bimodule $\{X,Y\}$ is \emph{good} if the graph $G$ with the labelling induced by $\{X,Y\}$ is \nlcrf.
The following proposition comes immediately from lemma~\ref{lem:nlcrfprime}.
\end{definition}

\begin{proposition}\label{prop:goodbimod}
$G$ is NLC-2 if and only if $G$ has a good $2$-bimodule.
\end{proposition}

\begin{lemma}\label{lem:lemstronggoodsplit}
If $G$ has a good $2$-bimodule $\{X,Y\}$ which is a split, then $G$ has a good $2$-bimodule which is a strong split.
\end{lemma}
\begin{proof}
There is a node $\alpha$ in the split decomposition tree and $\emptyset \subsetneq I\subsetneq \{1,\ldots ,d(\alpha )\}$ such that $\{X,Y\} = \{\cup_{i\in I} \compa{i}, \cup_{i\not \in I} \compa{i}\}$.
Let $l: V \to \{1,2\}$ be the labelling of $G$ induced by $\{X,Y\}$. For all $i\in \{1,\ldots ,d(\alpha)\}$, $(G[\compa{i}],\fnind{l}{\compa{i}})$ is \nlcrf (where $l\vert_W$ is the function $l$ restricted at $W$).

Let $l'$ be the $2$-labelling of $V$ such that for all $i$, and $v\in \compa{i}$, $l(v)=1$ if and only if $v$ has a neighbour outside of $\compa{i}$.
For all $i$, either $\fnind{l}{\compa{i}} = \fnind{l'}{\compa{i}}$, or $\forall v\in \compa{i}$, $l(v)=2$. Then for all $i$,  $(G[\compa{i}],\fnind{l'}{\compa{i}})$ is \nlcrf, and thus $(G,l')$ is \nlcrf.
Since there is a dominating vertex in the characteristic graph of $\alpha$, there is a $j$ such that the labelling induced by the strong split $\{\compa{j}, V\setminus \compa{j}\}$ is $l'$. Thus the strong split $\{\compa{j}, V\setminus \compa{j}\}$ is good.
\end{proof}

Previous lemma on $\overline{G}$ say that if $G$ has a good
$2$-bimodule $\{X,Y\}$ which is a co-split, then $G$ has a good
$2$-bimodule which is a strong co-split. The following lemma is similar to Lemma~\ref{lem:lemstronggoodsplit}.

\begin{lemma}\label{lem:lemstronggoodbijoin}
If $G$ has a good $2$-bimodule $\{X,Y\}$ which is a bi-join, then $G$ has a good $2$-bimodule which is a strong bi-join.
\end{lemma}

\begin{algorithm}[t]
\KwIn{A graph $G$} 
\KwResult{\kwyes iff $G$ is NLC-2}
$\ms \gets$ the set of strong splits, co-splits and bi-joins of $G$ \;
\ForEach{$\{X,Y\} \in \ms$}{
$l$ $\gets$ the labelling of $G$ induced by $\{X,Y\}$ \;
\lIf{$(G[X],G[Y],l)$ is \nlcrf}{\Return{\kwyes} \;}
}
\Return{\kwno} \;
\caption{Recognition of prime NLC-2 graphs}
\label{algo:recprime}
\end{algorithm}

\begin{theorem}
Algorithm~\ref{algo:recprime} recognises prime NLC-2 graphs, and its time complexity is $O(n^2 m)$.
\end{theorem}
\begin{proof}
Trivially if the algorithm return Yes, then $G$ is NLC-2. On the other hand, 
by proposition~\ref{prop:goodbimod}, and lemmas~\ref{lem:lemstronggoodsplit} and~\ref{lem:lemstronggoodbijoin}, if $G$ is NLC-2, then it has a good strong $2$-bimodule and the algorithm returns Yes.

The set $\ms$ can be computed 
using algorithms for 
computing split decomposition on $G$ and $\overline{G}$, and bi-join decomposition 
on $G$. Note that it is not required to use a linear time algorithm
for split decomposition~\cite{Dahlhaus00}: some simpler algorithms run
in $O(n^2
m)$~\cite{Cunningham82,Gabor89}.
\cite{DeMontgolfieR05,DeMontgolfieR05b} show that bi-join
decomposition can be computed in linear time, using a reduction to
modular decomposition. But there also,  modular decomposition
algorithms simpler than~\cite{McConnell99} may be used.
The set $\ms$ has $O(n)$ elements. Testing if a $2$-bimodule is good takes 
$O(nm)$ using algorithm~\ref{algo:treecomp}. So total running time is $O(n^2 m)$.
\end{proof}

\subsection{NLC-2 decomposition}

Using lemma~\ref{lem:nlcprime}, modular decomposition and algorithm~\ref{algo:recprime}, we get:
\begin{theorem}
NLC-2 graphs can be recognised in $O(n^2 m)$, and a NLC-2 expression can be generated in the same time.
\end{theorem}

\section{Graph isomorphism on NLC-2 graphs}\label{sectiso}

\subsection{Graph Isomorphism on  \nlcrf prime graphs}

The following propositions are direct consequences of properties (linear and degenerate) of \scuts.

\begin{proposition}
Consider a symmetric $S\in \bs$.
Two graphs $G$ and $H$ are isomorphic if and only if there is a bijection $\pi$ between $\mp_S(G)$ and $\mp_S(H)$ such that for all $P\in \mp_S(G)$, $G[P]$ is isomorphic to $H[\pi(P)]$. 
\end{proposition}

\begin{proposition}
Let a non-symmetric $S\in \bs$ and let $G$ and $H$ be two graphs.
Let $\mp'_S(G)= (P_1,\ldots , P_k)$ and $\mp'_S(H)= (P'_1,\ldots , P'_{k'})$  
then $G$ and $H$ are isomorphic if and only if $k=k'$ and for all $i\in \{1,\ldots , k\}$, $G[P_i]$ is isomorphic to $H[P'_i]$.
\end{proposition}

By the previous 2 propositions, two \nlcrf 2-labelled graphs $G$ and $H$ are isomorphic if and only if there is an isomorphism between their canonical \nlcrf decomposition tree which respects the order of children of \texttt{linear} nodes. 
This isomorphism can be tested in linear time, thus isomorphism of \nlcrf graphs can be done in $O(n m)$ time.

\mynote{fait citer une reference pour iso d'arbres ?}

\subsection{Graph isomorphism on prime NLC-2 graphs}

\begin{algorithm}[t]
%%\dontprintsemicolon 
\SetKw{KwOr}{or}
\SetKw{KwAnd}{and}
\SetKw{KwFail}{fail with}
\SetKw{KwCont}{continue}
\KwIn{Two prime NLC-2 graphs $G$ and $H$}
\KwResult{\kwyes if $G\iso H$, \kwno otherwise}
$\ms \gets$ the set of strong splits, co-splits and bi-joins of $G$ \;
$\ms' \gets$ the set of strong splits, co-splits and bi-joins of $H$ \;
\lIf{there is no good $2$-bimodule in $\ms$}{\KwFail ``$G$ is not NLC-2''}\;
$\{X,Y\} \gets$ a good $2$-bimodule in $\ms$ \;
$l$ $\gets$ the labelling of $G$ induced by $\{X,Y\}$ \;
\ForEach{$\{X',Y'\} \in \ms'$ such that $\{X',Y'\}$ is good}{
$l'$ $\gets$ the labelling of $H$ induced by $\{X',Y'\}$ \;
\If{$\vert X\vert >1$ \KwAnd $\vert Y \vert > 1$ \KwAnd $\{X,Y\}$ is a bi-join}{
  \lIf{$(G,l) \iso (H,l')$ \KwOr $(G,l) \iso (H,s(l'))$}{\Return{\kwyes \;}}
}\lElse{
  \lIf{$(G,l) \iso (H,l')$}{\Return{\kwyes \;}
  }
}
}
\Return{\kwno} \;
\caption{Isomorphism for prime NLC-2 graphs}
\label{algo:isoprime}
\end{algorithm}

\begin{theorem}
Algorithm~\ref{algo:isoprime} test isomorphism between two prime NLC-2 graphs in time $O(n^2 m)$.
\end{theorem}
\begin{proof}
If the algorithm returns ``yes'', then trivially $G\iso H$.
On the other hand suppose that $G\iso H$ and let $\pi: V(G) \to V(H)$ be a bijection such that $\{u,v\} \in E(G)$ iff $(\pi(u),\pi(v))\in E(H)$.
Then $\{X',Y'\}$ with $X'=\pi(X)$ and $Y'=\pi(Y)$ is a good $2$-bimodule if $H$. 
If $\min(\vert X \vert, \vert Y \vert)>1$ and $\{X',Y'\}$ is a bi-join, then by definition there is two labelling induced by $\{X,Y\}$, and $(G,l)\iso (H,l')$ or $(G,l) \iso (H,s(l'))$. Otherwise the labelling is unique and $(G,l)\iso (H,l')$.

The sets $\ms$ and $\ms'$ can be computed in $O(n^2)$ time using linear time algorithms for 
computing split decomposition on $G$ and $\overline{G}$, and bi-join decomposition on $G$. 
The sets $\ms$ and $\ms'$ have $O(n)$ elements. Test if a $2$-bimodule is good take $O(nm)$ 
using algorithm~\ref{algo:treecomp}, and test if two $2$-labelled prime graphs are isomorphic take also $O(nm)$. 
Thus the total running time is $O(n^2 m)$.
\end{proof}

\subsection{Graph isomorphism on NLC-2 graphs}

It is easy to show that graph isomorphism on prime NLC-2 graphs with an additional labels into $\{1,\ldots ,q\}$ can be done in $O(n^2 m)$ time. For that, we add the additional label of $v$ at the leaf corresponding to $v$ in the \nlcrf decomposition tree. 

We show that we can do graph isomorphism on NLC-2 graphs in time $O(n^2 m)$, using the modular decomposition and algorithm~\ref{algo:isoprime}.
Let $\mm(G)$ and $\mm(H)$ be the modular decomposition of $G$ and $H$. 
For $M\in \mm(G)$, let $G_M$ be $G[M]$, and for $M\in \mm(H)$, let $H_M$ be $H[M]$. 
Let $G^*_M$ be the characteristic graph of $G_M$ (note that $\vert V(G^*_M) \vert$ is the number of children of $M$ in the modular decomposition tree). 
Let $\mm_{(i,*)} = \{ M \in \mm(G) \cup \mm(H) : \vert M\vert =i\}$,
let $\mm_{(*,j)} = \{ M \in \mm(G) \cup \mm(H) : \vert V(G^*_M) \vert =j\}$ and
let $\mm_{(i,j)} = \mm_{(i,*)} \cap \mm_{(*,j)}$. 
Note that $\sum_{j=1}^n (\mm_{(*,j)} \times j)$ is the number of vertices in $G$ plus the number of edges in the modular decomposition tree, and thus is at most $3 n - 2$.

\begin{algorithm}[t]
\KwIn{Two NLC-2 graphs $G$ and $H$}
\KwResult{\kwyes if $G\iso H$, \kwno otherwise}
\lFor{every $M\in \mm(G) \cup \mm(H)$ such that $\vert M \vert =1$}{$l(M) \gets 1$} \;
\For{$i$ from $2$ to $n$}{
\For{$j$ from $2$ to $i$}{
Compute the partition $\mp$ of $\mm_{(i,j)}$ such that $M$ and $M'$ are in the same class of $\mp$ if and only if $(G^*_M,l) \iso (G^*_{M'},l)$. \;
\ForEach{$P\in \mp$}{
$a$ $\gets$ a new label (an integer not in $\operatorname{Img}(l)$) \;
For all $M\in P$, $l(M) \gets a$ \;
}
}
}
\caption{Isomorphism on NLC-2 graphs}
\label{algo:iso}
\end{algorithm}

\begin{theorem}
Algorithm~\ref{algo:iso} tests isomorphism between two NLC-2 graphs in time $O(n^2 m)$.
\end{theorem}
\begin{proof}
The correctness comes from the fact that at each step, for all $M, M' \in \mm(G) \cup \mm(H)$ such that $l(M)$ and $l(M')$ are set, $G_M$ and $G_{M'}$ are isomorphic if and only if $l(M) = l(M')$.
The total time $f(n,m)$ of this algorithm is $O(n^2 m)$ since (``big Oh'' is omitted):
\begin{eqnarray*}
&f(n,m) & \le \sum_i \sum_j \left (j^2 m \vert \mm_{(i,j)} \vert^2 \right ) \le m \sum_j \left (j^2 \sum_i \left (\vert \mm_{(i,j)} \vert^2 \right ) \right )\\
 &       & \le m \sum_j \left ( j^2 \vert \mm_{(*,j)} \vert^2 \right ) \le m \sum_j \left ( \left ( j \vert \mm_{(*,j)} \vert \right)^2 \right )  \le n^2 m.
\end{eqnarray*}
\vspace{-0.8cm}
\end{proof}

%%%%%%%%%%%%%%%%%%%%%%%%%%%%%%%%%%%%%%%%%%%%%%%%%%%%%%%%%%%%%%%%%%%%%%%%%%%%%%%%
{
\small

\bibliography{nlc2}
\bibliographystyle{plain}
}
%%%%%%%%%%%%%%%%%%%%%%%%%%%%%%%%%%%%%%%%%%%%%%%%%%%%%%%%%%%%%%%%%%%%%%%%%%%%%%%%
\newpage
\section*{Appendix}
\renewcommand{\thesection}{A}
\setcounter{subsection}{0}

\subsection{Proof of lemma~\ref{lem:vanherpe}}
\begin{quote}
Let $G = \biptri$ be a \bt such that every \btmodule has size $1$.
Let $(x_1,\ldots, x_{\vert X\vert})$ be $X$  sorted by $(d_j(x),d_m(x))$ in lexicographic decreasing order.
If $(A,B)$ is a \truc of $G$, then there is a $k\in \{0,\ldots , \vert X \vert\}$ such that $A \cap X =\{x_1,\ldots ,x_k\}$.
\end{quote}
\begin{proof}
For all $v\in A\cap X$, $d_j(v) \ge \vert B\cap Y \vert$, and 
for all $v\in B\cap X$, $d_j(v) \le \vert B\cap Y \vert$.
Moreover, if there is a $v\in B\cap X$ with $d_j(v) = \vert B\cap Y \vert$, then $d_m(v)=0$.
Let $C= \{ v\in X  : d_j(v) = \vert B\cap Y \vert \text{ and } d_m(v) = 0\}$.
Then $C$ is a \btmodule of $G$, and thus $\vert C \vert \le 1$.
Every vertex in $A\cap X\setminus C $ are before every vertex in $B\cap X \setminus C$ in the ordering. 
Moreover, if $\vert C \vert >0$, then vertices in $A\cap X\setminus C$ are before the vertex in $C$, and vertices in $B\cap X\setminus C$ are after the vertex in $C$ in the ordering.
\end{proof}

\subsection{Proof of lemma~\ref{lem:vanherpe2}}
\begin{quote}
Let $k\in \{0,\ldots, \vert X \vert\}$ and $k'\in \{0,\ldots, \vert Y \vert\}$.
Then $(A,(X\cup Y) \setminus A)$,
where $A=\{x_1,\ldots, x_k, y_1,\ldots, y_{k'} \}$,
is a \truc of $G$  if and only if 
$\sum_{i=1}^k d_j(x_i) - \sum_{i=1}^{k'} d_j(y_i) = k \times ( \vert Y \vert - k')$ and $\sum_{i=1}^k d_m(x_i) - \sum_{i=1}^{k'} d_m(y_i) = 0$.
\end{quote}
\begin{proof}
The ``If'' part is by definition. Now let us consider the ``Only if'' part. Let us assume that the degree
condition holds.
We will denote $a$ the number of join edges between 
$A\cap X$ and $B \cap Y$, $b$ the number of join edges between $A \cap X$ and $A\cap Y$, and $c$ the number of mixed edges between $A \cap X$ and $A\cap Y$.
Note that $a \le k (\vert Y\vert - k')$, $a+b= \sum_{i=1}^{k} d_j(x_i)$ and $b\le \sum_{i=1}^{k'} d_j(y_i)$, thus $a \ge k (\vert Y\vert - k')$. So we have $a = k (\vert Y\vert - k')$, and $\sum_{i=1}^{k'} d_j(y_i) - b =0$. In other words, there is only join edges between $A\cap X$ and $B \cap Y$, and there is no join edges between $A\cap Y$ and $B\cap X$.
Now since there is only join edges between $A\cap X$ and $B \cap Y$, $ c = \sum_{i=1}^k d_m(x_i) = \sum_{i=1}^{k'} d_m(y_i)$, thus there is no mixed edges between  $A\cap Y$ and $B\cap X$.
\end{proof}

\subsection{Algorithm to compute $\mp'_S$ when $S$ is non-symmetric}
\linesnumbered
\begin{algorithm}
\SetKw{KwOr}{or}
\SetKw{KwAnd}{and}
\KwIn{A $2$-labelled graph $G$, and a non-symmetric $S\subseteq\bs$} 
\KwOut{$\mp'_S$}
\SetKw{Vi}{$V_{i}$}
\Vi $\gets \{v : v \in V$ and $l(v)=i\}$ \;
\lIf{$(1,1) \in S$}{
  \nllabel{li:aa} $\setmod_1 \gets$ co-connected components of $G[V_1]$\; 
}
\lElse{
  \nllabel{li:ab} $\setmod_1  \gets$ connected components of $G[V_1]$\; 
}
%%%%%%%%
\lIf{$(2,2) \in S$}{
  \nllabel{li:ac}	
  $\setmod_2 \gets$ co-connected components of $G[V_2]$\; 
}
\lElse{
  \nllabel{li:ad} 
  $\setmod_2  \gets$ connected components of $G[V_2]$\; 
}

\nllabel{li:abiparti}$\mathcal{B}=(\setmod_1,\setmod_2,E_j,E_m)$ $\gets$ 
the bipartite trigraph between the elements of $\setmod_1$ and $\setmod_2$\ ;

\nllabel{li:ansstart}
$(x_1,\ldots, x_{\vert \setmod_1\vert})$ $\gets$ $\setmod_1$ sorted by lexicographic order on $(-d_j(v),-d_m(v))$ \;
$(y_1,\ldots, y_{\vert \setmod_2\vert})$ $\gets$ $\setmod_2$ sorted by lexicographic order on $(d_j(v),d_m(v))$ \;
$\mp' \gets ()$ ; 
$l \gets 0$; $l' \gets 0$; 
$k'\gets 0$ ; $k\gets 0$ \; 
$s_j \gets 0$ ; $s_m \gets 0$ ; $s'_j \gets 0$ ; $s'_m \gets 0$ \;
\While{$k \le \vert \setmod_1 \vert$}{
\While{\nllabel{li:nswhile2}$s_j - s'_j < k \times (\vert \setmod_2 \vert - k')$ \KwOr ( $s_j - s'_j = k \times (\vert \setmod_2 \vert - k')$ \KwAnd $s_m>s'_m$)}{
$k'\gets k'+1$ ; $s'_j \gets s'_j + d_j(y_{k'})$ ;  $s'_m \gets s'_m + d_m(y_{k'})$ \; 
}
\If{$s_j - s'_j = k \times (\vert \setmod_2 \vert - k')$ \KwAnd $s_m=s'_m$} {
  add $\{x_{l+1},\ldots, x_k\} \cup \{y_{l'+1}\ldots, y_{k'}\}$ at the end of $\mp'$ ;
  $l\gets k$ ; $l' \gets k'$ \;

\If{$s_j - s'_j - d_j(y_{k+1}) = k \times (\vert \setmod_2 \vert - k' -1)$ \KwAnd $s_m=s'_m + d_m(y_{k+1})$} {
  $k'\gets k'+1$ ; $s'_j \gets s'_j + d_j(y_{k'})$ ;  $s'_m \gets s'_m + d_m(y_{k'})$ \; 
  add $\{y_{k'}\}$ at the end of $\mp'$ ;
  $l' \gets k'$ \;
}
}
$k\gets k+1$ ; $s_j \gets s_j + d_j(x_k)$ ;  $s_m \gets s_m + d_m(x_k)$ \;
}
remove $\emptyset$ form $\mp'$, if any \;
\lIf{$(2,1) \in S$}{reverse $\mp'$}\; 
\Return{\nllabel{li:ansend}$\mp'$}

\caption{Computation of $\mp'_S$ when $S$ is non-symmetric}
\label{algo:mppcomp}
\end{algorithm}
\linesnotnumbered

\begin{proof}
{\bf Correctness:}
Algorithm~\ref{algo:mppcomp} generates all the \trucs of $\btcal$. At any time, $s_j=\sum_{i=1}^k d_j(x_i)$, $s_m=\sum_{i=1}^k d_m(x_i)$, $s'_j=\sum_{i=1}^{k'} d_j(y_i)$ and $s'_m=\sum_{i=1}^{k'} d_m(y_i)$.
In $\mathcal{B}$, every \btmodule has size $1$, otherwise there is a mono-coloured module in $G$ of size at least $2$.
If $(A,B)$ is a \truc, then by lemma~\ref{lem:vanherpe} on $(\setmod_1,\setmod_2,E_j,E_m)$ and $(\setmod_2,\setmod_1,E_j,E_m)$, there is a $a$ and $b$ such that $A\cap \setmod_1=\{x_1,\ldots, x_a\}$ and $A\cap \setmod_2 = \{y_1,\ldots, y_{b}\}$.
At any time, $(A',(\setmod_1\cup\setmod_2)\setminus A')$ with $A'=\{x_1,\ldots, x_{l}, y_1,\ldots, y_{l'}\}$ is the last \truc found.
At $k=a$, the {\bf while} line \ref{li:nswhile2} will stop when $s_j -
s'_j = k \times (\vert \setmod_2 \vert - k')$ since for every $v \in
A\cap \setmod_2$, $d_j(v)\le k$, and $s'_j + k \times (\vert \setmod_2
\vert - k')$ decrease with $k'$. Moreover, when the {\bf while} loop stops, $s_m = s'_m$ since $s'_m$ increase with $k'$. 
Thus if $b\ne k'$, then $\{y_{k'+1},\ldots y_b\}$ is a \btmodule and $b=k'+1$ (since every \btmodule has size $1$). In all cases the algorithm finds $(A,B)$, and adds the partition in $\mp'$.

{\bf Complexity}:
As we see in proof of theorem~\ref{th:correctness_mpcomp}, every instruction lines [\ref{li:aa}-\ref{li:abiparti}] can be done in linear time, and clearly every instruction lines [\ref{li:ansstart}-\ref{li:ansend}] can be done in linear time, thus the total running time is $O(n+m)$.
\end{proof}

%%%%%%%%%%%%%%%%%%%%%%%%%%%%%%%%%%%%%%%%%%%%%%%%%%%%%%%%%%%%%%%%%%%%%%%%%%%%%%%%

\end{document}